\documentclass[useAMS,usegraphicx,usenatbib]{mn2e}

\newcommand{\ngca}{$N_{GC}$}
\newcommand{\mbha}{$M_{\bullet}$}

\newcommand{\ngc}{$N_{GC}~$}
\newcommand{\mbh}{$M_{\bullet}~$}

\def\lta{\mathrel{\spose{\lower 3pt\hbox{$\mathchar"218$}}
     \raise 2.0pt\hbox{$\mathchar"13C$}}}
\def\gta{\mathrel{\spose{\lower 3pt\hbox{$\mathchar"218$}}
     \raise 2.0pt\hbox{$\mathchar"13E$}}}

\title[The Globular Cluster/Central Black Hole Connection in Galaxies]
{The Globular Cluster/Central Black Hole Connection in Galaxies}
\author[Gretchen L. H. Harris and William E. Harris] 
{Gretchen L. H. Harris$^{1}$\thanks{E-mail: glharris@astro.uwaterloo.ca;
 harris@physics.mcmaster.ca}
and William E. Harris$^{2}$ \\
$^{1}$Department of Physics and Astronomy, University of Waterloo, 
Waterloo ON N2L 3G1, Canada\\
$^{2}$Department of Physics and Astronomy, McMaster University, Hamilton ON L8S 4M1, Canada}
\begin{document}

\date{Accepted  Received; in original form }

\pagerange{\pageref{firstpage}--\pageref{lastpage}} \pubyear{2002}

\maketitle

\label{firstpage}

\begin{abstract}
We explore the relation between the total globular population in 
a galaxy (\ngca) and the mass of its central black hole (\mbha).
Using a sample of 33 galaxies, twice as large as the original
sample discussed by Burkert \& Tremaine (2010), we find that
\ngc for elliptical and spiral galaxies
increases in almost precisely direct proportion to \mbha.
The S0-type galaxies by contrast do not follow a clear trend,
showing large scatter in \mbh at a given \ngca.  
After accounting for observational measurement uncertainty, we find
that the mean relation defined by the E and S galaxies
must also have an intrinsic or ``cosmic'' scatter
of $\pm 0.2$ in either log \ngc or log \mbha.  The residuals from this
correlation show no trend with globular cluster specific frequency.
We suggest that these two types of galaxy subsystems (central black hole
and globular cluster system) may be closely correlated because they
both originated at high redshift during the main epoch of hierarchical
merging, and both require extremely high-density conditions for
formation.
Lastly, we note that roughly $10\%$ of the galaxies in our sample 
(one E, one S, two S0) 
deviate strongly from the main trend, all in the sense that their \mbh
is at least $10\times$ smaller than would be predicted by the
mean relation. 
\end{abstract}


\section{Introduction}

Recently \citet{b1} (hereafter BT) 
have presented evidence that the total population of globular clusters 
in a galaxy is almost exactly proportional to the mass of its central 
supermassive black hole.
In rough terms such a relationship may not, on the surface, be terribly surprising because 
(as BT discuss) 
\ngc is nearly linearly related to the
luminosity or total mass of the host galaxy (e.g. Harris 1991),
and \mbh scales nearly linearly with the galaxy bulge mass
\citep{w3, f2}.
At a more detailed level, however, the behavior of the GC population
with galaxy size and type is much more complex \citep{b2, p1}, 
and it is not obvious that the
\ngc-\mbh correlation should be as tight
and strongly defined as it is -- as BT discuss, more so than is
 warranted by the measurement uncertainties.
A potentially more surprising result they find from their sample is that the
total \emph{mass} contained in the globular cluster system is nearly equal
to that of the central black hole.

In this paper, we investigate the \ngc vs. \mbh correlation further and
discuss additional implications arising from it.

\section[]{The Database}

The BT result was based on just 16 galaxies: 11 ellipticals (E), 
3 spirals (S), and 2 lenticulars (S0), and only 13 were used in 
deriving the numerical relation between total cluster population
and central black hole mass.  Given that the number of galaxies 
with either measured \ngc or \mbh is considerably larger than this, we chose to try 
and add to the sample.  We carried out an independent search of
 the literature using galaxies included 
in the the \mbh databases of \citet{g8}, \citet{g7}, and \citet{g4}, 
whose globular cluster systems (GCSs) had also been studied and for which it was possible 
to make reasonable estimates of \ngca.  As a 
result we now have a database of 33 galaxies: 
21 E, 4 S, and 8 S0, doubling the BT sample. 
Many other galaxies have published measurements of \emph{either} one or the other
(more than 70 with \mbh and well over 200 for \ngca), but 
these two types of observational studies have been carried out originally
for very different purposes, so
it is perhaps not suprising that the overlap
between them is as yet quite limited.

In Table 1 we list the results of our compilation.  Successive columns give
the galaxy name, galaxy type, absolute $V$ magnitude, total cluster 
population \ngc and data source, specific frequency (number of clusters
per unit galaxy luminosity), black hole mass \mbh and source.
References for the \ngc and \mbh values are provided in the footnote.  
For a few of the galaxies in this list,
 the original GC study quoted only a value for the number of directly
observed GCs and not a total cluster
population.  For these we reconstructed a total, correcting for spatial and
magnitude incompleteness with standard methods described, for example,
in \citet{h2}.  
In those cases we reference both the original GC study and this paper. 
In some of these cases the resulting
\ngc value we believe to be accurate to only a factor of two, but these are still
quite useful for the purposes of this discussion; and it should be noted
that very similar uncertainties also
apply to many of the black-hole masses.  
For a thorough recent review of the current \mbh database and analysis issues, see 
\citet{g8}.

\begin{table*}
\caption{SMBH Masses and Globular Cluster Numbers}
\label{alldata}
\begin{tabular}{lccccccc}
\hline
Galaxy & Type & $M_V$ & $N_{GC} (\pm)$ & Source & $S_N$ & $M_{BH}/M_{\odot} (\pm)$ & Source \\
\\
NGC221 (M32) & E2 & $-19.4$ & $<10$ & 1,2 & $<0.2$ & $3.4(+0.6,-0.6) \times 10^6$ & 3 \\
NGC821 & E4 & $-21.0$ & 320(45) & 4 & 1.3 & $4.2(+2.8,-0.8) \times 10^7$ & 5 \\
NGC1399 & E1 & $-22.3$ & 6625(1180) & 6,7 & 7.6 & $9.0(+0.9,-1.0) \times 10^8$ & 8,9 \\
NGC2778 & E2 & $-19.6$ & 50(30) & 1,10 & 0.7 & $1.6(+0.9,-0.02) \times 10^7$ & 5 \\
NGC3377 & E5 & $-20.1$ & 266(66) & 11 & 2.4 & $1.1(+1.1,-0.1) \times 10^8$ & 5 \\
NGC3379 (M105) & E1 & $-20.9$ & 270(68) & 12 & 1.2 & $1.2(+0.8,-0.6) \times 10^8$ & 13 \\
NGC4261 & E2 & $-22.7$ & 530(100) & 1,14 & 0.4 & $5.5(+1.1,-1.2) \times 10^8$ & 15 \\
NGC4374 (M84) & E1 & $-22.3$ & 4300(1200) & 16 & 5.2 & $1.5(+1.1,-0.6) \times 10^9$ & 17 \\
NGC4472 (M49) & E2 & $-23.0$ & 7800(850) & 16 &  4.9 & $1.5(+0.6,-0.6) \times 10^9$ & 18 \\
NGC4473 & E5 & $-20.7$ & 376(97) & 16 & 2.0 & $1.3(+0.5,-0.94) \times 10^8$ & 5 \\
NGC4486 (M87) & E1 & $-22.7$ & 13300(2000) & 16,19 & 11.1 & $6.4(+0.5,-0.5) \times 10^9$ & 20 \\
NGC4486A & E2 & $-18.7$ & 11(11) & 16 & 0.4 & $1.3(+0.5,-0.4) \times 10^7$ & 21 \\
NGC4486B & E0 & $-17.7$ & 4(11) & 16 & 0.4 & $<1.0 \times 10^9$ & 22 \\
NGC4552 (M89) & E0 & $-21.4$ & 1200(250) & 16,19 & 3.3 & $4.8(+0.8,-0.8) \times 10^8$ & 23 \\
NGC4564 & E6 & $-20.0$ & 213(31) & 16 & 2.1 & $6.9(+0.4,-1.0) \times 10^7$ & 5 \\
NGC4621 (M59) & E5 & $-21.3$ & 800(355) & 16 & 2.4 & $4.0(+0.6,-0.6) \times 10^8$ & 23 \\
NGC4649 (M60) & E2 & $-22.4$ & 4745(1100) & 16 & 5.2 & $4.5(+1.0,-1.0) \times 10^9$ & 5 \\
NGC4697 & E6 & $-21.0$ & 229(50) & 24 & 0.9 & $2.0(+0.2,-0.2) \times 10^8$ & 5 \\
NGC5128 & Ep & $-22.1$ & 1300(300) & 25 & 1.9 & $7.0(+1.3,-3.8) \times 10^7$ & 26 \\
NGC5813 & E2 & $-22.1$ & 1650(400) & 27 & 2.4 & $7.0(+1.1,-1,1) \times 10^8$ & 23 \\
NGC5846 & E0 & $-22.6$ & 4700(1200) & 28 & 4.3 & $1.1(+0.2,-0.2) \times 10^9$ & 23 \\
NGC1023 & SB0 & $-21.3$ & 221(100) & 29 & 0.7 & $4.6(+0.5,-0.5) \times 10^7$ & 30 \\
NGC1316 & S0 & $-22.8$ & 647(100) & 31 & 0.5 & $1.5(+0.8,-0.8) \times 10^8$ & 32 \\
NGC3115 & S0 & $-21.1$ & 550(150) & 33 & 2.0 & $9.6(+5.4,-2.9) \times 10^8$ & 34,35 \\
NGC3585 & S0 & $-21.8$ & 300(100) & 1,36 & 0.6 & $3.4(+1.5,-0.6) \times 10^8$ & 37 \\
NGC4350 & S0 & $-20.1$ & 196(60) & 16 & 1.8 & $6.0(+1.5,-0.9) \times 10^8$ & 38 \\
NGC4382 & S0 & $-22.0$ & 1100(181) & 16 & 1.3 & $1.3(+5.2, -1.2) \times 10^7$ & 39 \\
NGC4459 & S0 & $-20.3$ & 218(28) & 16 & 1.7 & $7.4(+1.4,-1.4) \times 10^7$ & 40 \\
NGC7457 & S0 & $-19.6$ & 178(75) & 41 & 2.6 & $4.1(+1.1,-1.7) \times 10^6$ & 5  \\
Milky Way & Sbc & $-20.6$ & 160(20) & 42 & 0.9 & $4.1(+0.4,-0.4) \times 10^6$ & 43,44 \\
NGC224 (M31) & Sb & $-21.7$ & 450(100) & 45 & 0.9 & $1.4(+0.9,-0.3) \times 10^8$ & 46 \\
NGC3031 (M81) & Sb & $-21.5$ & 172(100) & 47 & 0.4 & $8.0(+2.0,-1.1) \times 10^7$ & 48 \\
NGC4594 (M104) & Sa & $-22.4$ & 1900(300) & 4 & 2.1 & $5.5(+5.3,-4.0) \times 10^8$ & 49 \\
\hline
\end{tabular}

Sources:  (1) This paper, (2) \citet{l1}, (3) \citet{v1}, (4) \citet{s1}, 
(5) \citet{g3}, (6) \citet{b3}, (7) \citet{d1}, (8) \citet{g5}, (9) \citet{h3},
(10) \citet{g1}, (11) \citet{k1}, (12) \citet{h2}, (13) \citet{g13},
(14) \citet{g9}, (15) \citet{f1}, (16) \citet{p1}, (17) \citet{b4},
(18) \citet{s2}, (19) \citet{t2}, (20) \citet{m3}, (21) \citet{n1},
(22) \citet{g8}, (23) \citet{g7}, (24) \citet{j1}, (25) \citet{h1},
(26) \citet{c1}, (27) \citet{h4}, (28) \citet{f3}, (29) \citet{l2},
(30) \citet{b5} (31) \citet{v3}, (32) \citet{n2}, (33) \citet{h6},
(34) \citet{e1}, (35) \citet{t1}, (36) \citet{p3}, (37) \citet{g14},
(38) \citet{p4}, (39) \citet{g10} (40) \citet{s3}, (41) \citet{c2},
(42) \citet{h5}, (43) \citet{g11}, (44) \citet{g12}, (45) \citet{b6},
(46) \citet{b7}, (47) \citet{s4}, (48) \citet{d2}, (49) \citet{k2}
\end{table*}

The key plot of log \ngc vs. log \mbh for all galaxies in Table 1 is shown in Figure 1. 
If the two quantities are exactly proportional then the mean relation in this
log-log graph will have unit slope, and the data do follow a trend very close
to that.  Thus
the first conclusion to be drawn from this larger dataset 
is that we verify the essential result of BT, now extending over a
factor of $10^3$ in both quantities.
The main spine of the proportionality is defined primarily by the E galaxies 
which make up more than half the sample.  

Though a trend clearly exists, it is also true that some galaxies lie well off 
the range of scatter defined by the majority.  
Particularly disparate are the points for NGC 5128, 
the Milky Way (MW), and the S0s NGC 4382 and  NGC 7457. All have black hole
masses which are at least an order of magnitude too small for their cluster populations
when compared to the main trend.  For the MW, both \mbh and \ngc are very well known, and
we cannot arbitrarily choose to treat the cluster population in any different way from the other
galaxies in the list (for example by counting only the ``bulge'' GCs as BT suggest). 
In fact, given that clusters in the central bulge are generally more difficult to detect in external 
galaxies, a more reasonable `correction' to the MW sample would be to eliminate 
some fraction of the bulge clusters.  But since these make up
$\sim 30 \%$ of the total population at most, any such correction would still leave
the MW as an outlier in Figure 1.

 For NGC 5128, the cluster population estimates have consistently been in the range of 
1000-2000 and the most recent determination of 1300 \citep{h1} is now more soundly 
based than the GC totals for many of the other systems in Table 1.
To force this galaxy back to the mean line, either \ngc would need to be smaller by 
a factor of at least 3 (an impossibility 
since more than 600 normal GCs have been individually found and confirmed; see 
\citet{w2}); or the BH mass would have to be far larger (also now unlikely, as 
mentioned above).  The dust lane has 
made determination of \mbh rather challenging, but it does seem that the mass used here is
one of high confidence \citep{c1}.  Lastly, in the cases of NGC 4382 and NGC 7457 the BH 
masses are solid upper limits \citep{g6} 
and their clusters would essentially all have to disappear to make these galaxies 
fall anywhere near the main line.  

\begin{figure}
 \includegraphics[width=0.45\textwidth]{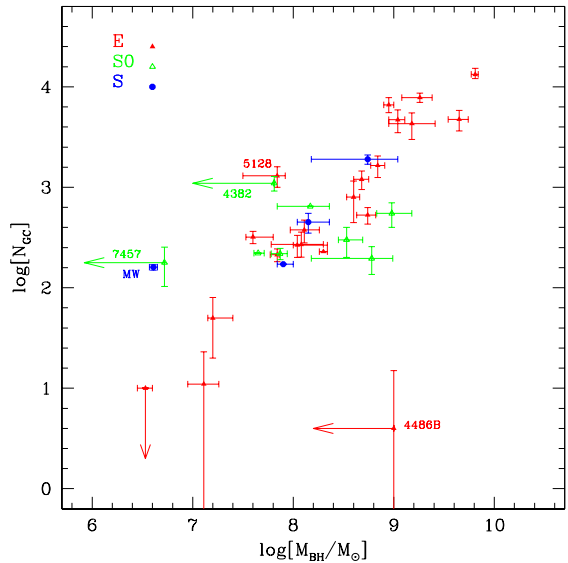}
 \caption{log \ngc is plotted against log \mbh for all 33 galaxies in our sample.  E galaxies are 
filled red triangles, S galaxies are filled blue circles, and open green triangles are S0
galaxies. For three galaxies only upper \mbh limits are measured and these are shown with the
horizontal arrows; for one we have only an upper limit on \ngca, shown as a vertical arrow.
The five galaxies which deviate most strongly from the main trend are labeled; see further
discussion in the text.}  
\end{figure}

Two other objects worth special comment are the compact ellipticals M32 and NGC 4486B 
which have vanishingly small GC populations.
Both are satellites of much larger systems (M31 and M87 respectively) and it has often
been suggested that the GCs they might initially have had were removed by tidal
stripping (for the outlying ones) or dynamical friction (for the ones in the core).
For M32, we decided to make a fresh estimate of the number of clusters it might have
by searching in the M31 cluster catalog of \citet{l1}.  There are 9 clusters
within a projected distance of 3 kpc of M32 \emph{and} within $\pm 100$ km/s of its
radial velocity.  From the same database, to within the count statistics
we find the same number of 
number of M31 clusters per unit area,  within that velocity range and at the
mean radial distance of M32; 
therefore we conclude that M32 has no statistically significant GC `signal'.
We adopt a conservative upper limit \ngc $< 10$.
For NGC 4486B, the GC population as measured in the comprehensive 
deep Virgo Cluster Survey \citep{p1} is indistinguishable from zero,
and the BH mass measurement is also only an upper limit, so it is not clear whether
or not this galaxy falls off the mean relation.  
In any case, we do not use these two small
ellipticals in the following analysis.

\section{Analysis}

Given that the BT results were based on a sample dominated by E galaxies, which make up
more than half of our list, we decided to look at the data sorted by subtype before trying
to find numerical correlations.  In Figure 2 we plot log \ngc vs. log \mbh as before, but
with E, S, and S0 galaxies in different panels.  A brief glance is enough to see that
the distributions show certain differences for the three galaxy types:
\begin{itemize}
\item The E galaxies span a range of three orders of magnitude
in both \mbh and \ngca, larger than for the S0 or S types,
and they show a strong correlation between log \mbh and log \ngc over that range,
 even in the low \mbh - \ngc part of the graph where there are only a few galaxies.
\item The four S galaxies cover a factor of only $\sim$10 
in \ngc and $\sim$100 in \mbha.  But, with the notable exception of the MW,
 the spirals appear to follow a trend similar to the ellipticals.
\item The S0 galaxies are quite different, occupying a similar \ngc range to
the spirals, but a much larger one in central black hole mass.   
No clear trend is present.
Arbitrarily increasing the central black hole mass  for the two S0s with 
log\mbh $\la$7 
would place them closer to the main group but as already
discussed above, the quoted values 
of \mbh are upper limits for both \citep{g6}.  
That S0s have rather small cluster populations (that is, low specific frequencies) relative to ellipticals
has long been known and can be illustrated well by 
examining the homogeneous ACS Virgo Cluster Survey lists of Peng et al. (2008).
Their sample contains 31 
galaxies identified as S0: more than half have \ngc $\la$ 100 and all 
but two are likely   
to have $<$ 400 clusters. 
\end{itemize}

\begin{figure*}
 \includegraphics[trim = 0mm 0mm 0mm 95mm, clip,width=0.90\textwidth]{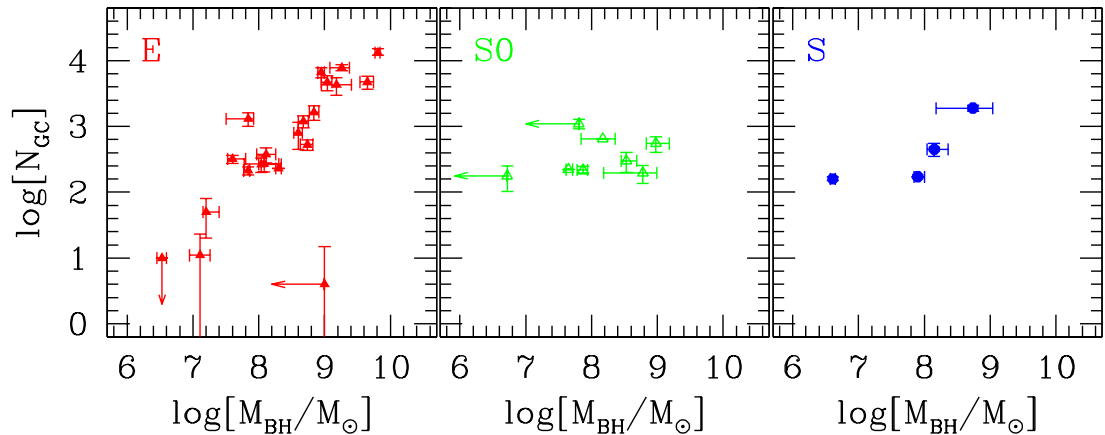}
 \caption{log \ngc versus log \mbh for the three types of galaxies (E, S0, S).} 
\end{figure*}

Because the E galaxies as a subclass define by far the most well determined correlation,
we use them to define the baseline relation.  We solved for 
a best-fit relation using the $\chi^2$ estimator described by \citet{t1}
which accounts for the observational measurement uncertainties in
both quantities for each galaxy individually.  The 18 ellipticals (excluding only
M32, NGC 4486B, and NGC 5128) yield a solution
\begin{equation}
{\rm log} N_{GC} \, = \, 
(-5.78 \pm 0.85) \, + \, (1.02 \pm 0.10) {\rm log} {M_{BH} \over {M_{\odot}}} 
\end{equation}
or, equivalently, 
\begin{equation}
{\rm log} {M_{BH}\over {M_{\odot}}} \, = \, 
(-5.66 \pm 0.29) \, + \, (0.98 \pm 0.10) {\rm log} {N_{GC}}  
\end{equation}
Including NGC 5128
does not change the best-fit relation significantly though it does increase 
the $\chi^2(min)$.

In short, from this larger database we find (as did BT)
that \emph{the mass of the central black hole scales nearly linearly with the globular
cluster population in the galaxy}.  If we force the  
slope in Equation (1) to be exactly 1.00,
then the best-fit relation would become
\begin{equation}
M_{BH}/M_{\odot} \, = \, 4.07 \times 10^5 \, N_{GC}
\end{equation}
or, expressed as a rule of thumb, about 250 globular clusters 
per $10^8 M_{\odot}$ black hole.

Figure 3 shows both the best-fit relation of Equation (1) and the unit-slope
relation, superimposed on the separate plots for the three types of host galaxies.
Plotting both of these llustrates that we have
no compelling reasons to reject an exact 1:1 relation.   It should be possible
to improve the fit with the addition of more galaxies, particularly those with log \mbh
$< 8$.  

\begin{figure*}
 \includegraphics[trim = 0mm 0mm 0mm 95mm, clip,width=0.90\textwidth]{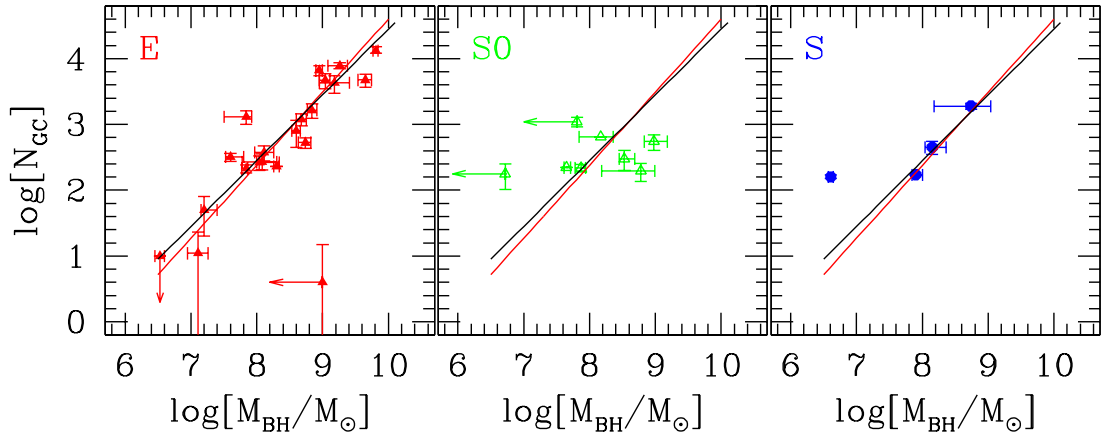}
 \caption{The same plots as in the previous figure, now with the best-fit relation
for the E galaxies plotted in all three panels (red line).  The \emph{black line}
in each panel is the relation with the forced assumption that the slope is
unity.}
\end{figure*}

The residual scatter about the mean (as indicated by the raw $\chi^2(min)$) 
is noticeably larger than the random measurement errors would nominally permit.
We have no particular reason to assume that the various carefully done studies
in the literature have systematically underestimated the measurement uncertainties in either
quantity, so we suggest that a real cosmic dispersion is also present.
To bring the formal $\chi^2$  per degree of freedom down to 1.0
at the best-fit point, we find that we need to introduce 
an additional \emph{intrinsic scatter} of
$\pm0.2$ in either log \mbh or log \ngca, or in linear terms a factor of $\simeq 1.6$.

For a normal Gaussian-type globular cluster luminosity function such
as in the Milky Way, with a mean at
$M_V = -7.3$ and dispersion $\sigma(M_V) = 1.3$ (e.g. Brodie \& Strader 2006), the average
cluster luminosity is $L_V = 1.2 \times 10^5 L_{\odot}$.  With a mean
$(M/L_V) = 2$ (e.g. \citet{m1}) the mean GC mass is then $2.4 \times 10^5 M_{\odot}$,
and the E-galaxy correlation becomes very roughly $M_{BH} \sim 1.5 M_{GC}(tot)$.
To well within a factor of two the total masses in these substructures are the same.

The spiral galaxies, with the exception of the MW as 
discussed above, fall extremely close to the E galaxy line.  In fact, adding
the three `good' spirals to the E-galaxy list and recalculating the solution
leads to no difference from the relation given in 
Equation (1).  The S0s are a different matter however;  while
a few fall nicely on or near the relationship, there are an equal 
number for which the \mbh - \ngc relation appears to be irrelevant.

It has long been known (beginning with {\citet{h2}) that the
GC specific frequency $S_N$, i.e. the number of clusters
per unit galaxy luminosity, varies with both galaxy type and
environment (see also {\citet{b2, p1}).  Since the galaxies in the list of
Table I have a large range in $S_N$, we thought it would be worthwhile
to ask whether any extra residual correlation of \mbh with $S_N$ 
might be present.  For example, consider two elliptical galaxies with the same
total \ngc but different luminosities.  By definition, the higher luminosity 
one must then have a lower $S_N$.  But, we might also expect the 
higher-luminosity galaxy to have a higher mass \mbh if the central black
hole mass is driven primarily by the host galaxy's mass.  If this
argument were valid, we should then see the opposite effect - i.e. 
an anticorrelation of \mbh with $S_N$.  In Figure \ref{snresid} we show
the residuals in \mbh from the mean relation of Equation (1) 
plotted as a function of $S_N$.  Interestingly no trend is
apparent, or at the very least, any such trend is obscured by the 
measurement scatter in \mbh and the small number of points at high $S_N$.
Of the two largest $S_N$ galaxies, one (M87) has a nearly 
``normal" \mbh and the other (NGC 1399) has a slightly lower than ``normal"
\mbha.  In short, from the present data we see no extra correlation
of central black hole mass with specific frequency.


The most sparsely populated area of the correlation is at low mass.
For \mbh much less than $10^8 M_{\odot}$, the presence of nuclear star
clusters (NCs) may also become relevant \citep{gra09}.  Both BH and NC
are part of the more general class of Central Massive Objects
\citep[CMO; see][]{w3, f2} and, as \citet{gra09} show, for 
some galaxies both are present at once.  For intermediate-mass galaxies
there is a gradual mass transition zone such that for \mbh $\la 
2 \times 10^7 M_{\odot}$ the NC mass tends to become dominant.
Inspecting Table 1 of Graham \& Spitler, we note that for the
MW and NGC 7457 particularly, adding the NC mass to \mbh would bring
them much closer to the mean E/S relation.  For several other 
galaxies, however (NGC 1023, 1399, 3115, 4697), no important change
would result.  Additional data may help clarify whether or not the NC mass
is an essential part of the picture.

As said earlier, the basic fact that the GC populations and central black hole
mass should be correlated is not surprising, because in rough terms bigger galaxies should
have bigger subsystems of all types.  What is more intriguing is that 
for the E and S galaxies especially, the correlation is so tight and
so nearly linear.  BT suggest that mergers may be responsible for the
growth of both these subcomponents of the galaxy.   However, in a more
general sense mergers will promote the growth of \emph{all} subcomponents
of the galaxy.
We speculate that a more relevant link between black holes and
globular clusters may be their age.  Both the central BH and the
majority of the GCs had their origins at high and nearly similar redshift,
during the major stages of hierarchical merging.
Most GCs have ages in the range $10 - 13$ Gyr, corresponding to
redshifts $z \sim 2 - 7$ \citep{m2, p2, w1, m4}.
For large galaxies the seeds of the central
black holes are in place before $z \sim 6$ and grow by further gas accretion
till $z \sim 2$ and later 
\citep[e.g.][among many others]{a1, h7, vol03, t3, v2, w4, hec04, sil08, mer10, kel10}.
Both types of structures clearly require formation conditions of 
extremely high gas density ($\ga 10^5 M_{\odot}$/pc$^3$
within small, parsec-sized volumes), which would have been easiest
to generate in large numbers at very early times when gas cloud
collisions and hierarchical merging events
were frequent and energetic \citep[cf. the discussions by][]{p1, r1}.  
In this scenario, \ngc and \mbh should be closely correlated simply because
they are both byproducts of similar extreme conditions 
at high-density locations during the main period of galaxy formation.

A less certain issue would be whether or not there is
a direct causal link between them.  For example, 
the early energy output from the dense, violent star formation conditions
in the central regions of a large galaxy that might have accompanied
the buildup of the massive BH 
might also have helped stimulate the formation of dense, massive
star clusters throughout the bulge and halo regions 
\citep[e.g.][]{h9,tan99,kra02,spr05,kau07}.
 
The rough equality of total \emph{mass} in the two subsystems
as shown above may, however, be largely a coincidence.  The reason is
that it applies only to the \emph{present} stage of evolution of both
subsystems.
The total mass in the globular cluster system has been
continually decreasing since their formation epoch:  during the
first several $\sim 10^7$ years of a young massive star cluster's life, it sheds a high
fraction of its initial protocluster gas due to SNe and stellar winds,
and the highest-mass stars evolve and disappear.  Over
the subsequent Hubble time, 
tidal stripping and evaporation remove most low-mass clusters,
and the higher-mass ones are significantly eroded \citep[see][for a review]{ves10}.
 These factors combined mean that the initial
mass in the GC system should have been many times larger than it is now.
By contrast, the BH mass has been continually growing over the same time
through gas accretion and stellar capture, though its most rapid growth
stage would have been at early times (cf.~the references cited above).
In other words, in their early stages the total mass in the GC population
must have been much larger than the initial BH mass (an order of
magnitude or more).  It is only now, after 
a Hubble time's worth of evolution in opposite senses, that their total masses are nearly
crossing.

The case of the S0 galaxies is extremely puzzling.  There seems to be no
obvious way in which the data for these systems is more uncertain than for
the E and S galaxies, leaving us in the position of questioning the sample 
and/or classification.  As already discussed, the two most discordant 
S0s have black hole masses which are upper limits and there appears to be no obvious
way in which they can be moved to more closely agree with the majority of the
galaxies in our sample.  Interestingly, they deviate from the norm in the same 
sense as the MW and NGC 5128 (discussed above).  The four galaxies which
are the most discordant then (1 E, 1, S and 2 S0 - roughly 
$10\%$ of our sample) all deviate in the sense 
that their \mbh values are too small compared with \ngca; and without the two
discordant S0s, the S0 sample overall becomes less peculiar.  However, all four 
"discordant" galaxies 
have normal specific frequencies for their galaxy type - i.e. \ngc
 is normal when compared to the luminosity/mass of the whole galaxy.
This suggests to us that the formation of GCs is more strongly correlated with overall
galaxy properties, and that in some cases \mbh is 
a later or less rapid development.

\begin{figure}
 \includegraphics[width=0.45\textwidth]{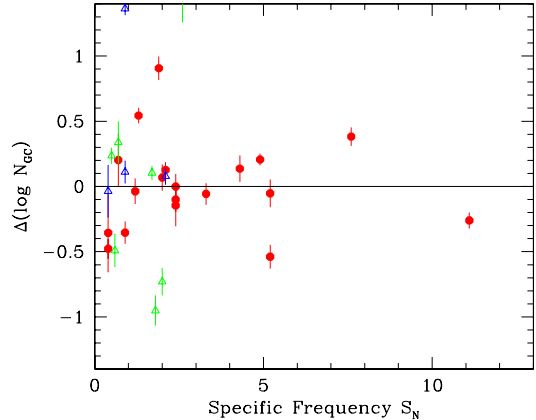}
 \caption{Residuals in \mbh from the mean relation in Equation (1) are plotted
versus globular cluster specific frequency.  E galaxies are shown as solid
red dots, S galaxies as blue triangles, and S0 galaxies as 
green triangles.  
} 
\label{snresid} 
\end{figure}

\section{Conclusions}

\begin{enumerate}
\item We have investigated the correlation 
between central black hole mass in a galaxy and the total globular 
cluster population suggested by \citet{b1}.
By doubling the sample of galaxies and reconstructing
the numbers for \ngc and \mbh from the primary literature, we find
results that confirm the basic trend they presented.  
\item Over three orders of magnitude, the elliptical and spiral galaxies
in the sample define a relation whereby \ngc $\sim$ $M_{BH}^{1.02 \pm 0.10}$,
a nearly linear proportionality.  By contrast, the S0 galaxies in the sample
do not follow the same trend, showing much larger scatter in \mbha.
\item Of 33 galaxies in the current sample, three or four 
(i.e. $\sim 10\%$) fall far enough
from the mean relation to be considered genuinely anomalous.  The Milky Way
is one of these, with a central BH mass far too small for its cluster
population.
\item We suggest that the source of this correlation is connected with
the epoch of origin of these structures.  Both the
 central BH and the bulk of the GC population had
their origins at high redshift ($z \sim 2-7$), both requiring extremely
high gas density conditions of $10^5 M_{\odot} {\rm pc}^{-3}$ and above.
\end{enumerate}

On observational grounds there is much room for improvement to
explore this intriguing correlation further with bigger samples.
Galaxies with already-known BH masses can be imaged with deep wide-field
photometry to determine their cluster populations, while the cores of many nearby
galaxies with well determined \ngc values can be investigated to
determine new BH masses. Some dozens of new datapoints can
be added to the current list. Larger samples should also allow
exploration of wider ranges of galaxy environment and host galaxy type,
and perhaps allow us to uncover more direct causal links between
these old substructures.

\section*{Acknowledgments}

We thank Nadine Neumayer, Karl Gebhardt, and Andreas Burkert for helpful discussions.  
The majority of this work was undertaken during a visit to ESO 
in Garching, sponsored by the ESO visitor programme. WEH also acknowledges financial 
support through a grant from the Natural Sciences and Engineering Research 
Council of Canada.

\label{lastpage}

\end{document}